\documentclass[aps,prl,superscriptaddress,twocolumn,showpacs,amsmath,amssymb,floats,]{revtex4}

\usepackage{txfonts}
\usepackage{amssymb}
\usepackage{graphicx}

\begin{document}

\title{Momentum-resolved measurement of electronic nematic susceptibility in the FeSe$_{0.9}$S$_{0.1}$ superconductor}

\author{C. Cai}
\affiliation{International Center for Quantum Materials, School of Physics, Peking University, Beijing 100871, China}

\author{T. T. Han}
\affiliation{International Center for Quantum Materials, School of Physics, Peking University, Beijing 100871, China}

\author{Z. G. Wang}
\affiliation{International Center for Quantum Materials, School of Physics, Peking University, Beijing 100871, China}

\author{L. Chen}
\affiliation{International Center for Quantum Materials, School of Physics, Peking University, Beijing 100871, China}

\author{Y. D. Wang}
\affiliation{International Center for Quantum Materials, School of Physics, Peking University, Beijing 100871, China}

\author{Z. M. Xin}
\affiliation{International Center for Quantum Materials, School of Physics, Peking University, Beijing 100871, China}

\author{M. W. Ma}
\affiliation{International Center for Quantum Materials, School of Physics, Peking University, Beijing 100871, China}

\author{Yuan Li}
\affiliation{International Center for Quantum Materials, School of Physics, Peking University, Beijing 100871, China}
\affiliation{Collaborative Innovation Center of Quantum Matter, Beijing 100871, China}

\author{Y. Zhang}\email{yzhang85@pku.edu.cn}
\affiliation{International Center for Quantum Materials, School of Physics, Peking University, Beijing 100871, China}
\affiliation{Collaborative Innovation Center of Quantum Matter, Beijing 100871, China}

\date{\today}

\begin{abstract}
Unveiling the driving force for a phase transition is normally difficult when multiple degrees of freedom are strongly coupled. One example is the nematic phase transition in iron-based superconductors. Its mechanism remains controversial due to a complex intertwining among different degrees of freedom. In this paper, we report a method for measuring the nematic susceptibly of FeSe$_{0.9}$S$_{0.1}$ using angle-resolved photoemission spectroscopy (ARPES) and an $in$-$situ$ strain-tuning device. The nematic susceptibility is characterized as an energy shift of band induced by a tunable uniaxial strain. We found that the temperature-dependence of the nematic susceptibility is strongly momentum dependent. As the temperature approaches the nematic transition temperature from the high temperature side, the nematic susceptibility remains weak at the Brillouin zone center while showing divergent behavior at the Brillouin zone corner. Our results highlight the complexity of the nematic order parameter in the momentum space, which provides crucial clues to the driving mechanism of the nematic phase transition. Our experimental method which can directly probe the electronic susceptibly in the momentum space provides a new way to study the complex phase transitions in various materials.

\end{abstract}


\maketitle

Nematic phase, a symmetry breaking phase that is translational invariant but breaks rotational symmetry, attracts great interests recently as it is tightly interwoven with high-$T_c$ superconductivity in both cuprates and iron-based superconductors \cite{r1,r2,r3,r4,r5,r6,r7}.  Therefore, it is important to understand the nematic phase transition, whose driving mechanism is closely related to the pairing interaction of high-$T_c$ superconductivity. For iron-based superconductors, the nematic phase transition occurs when the rotational symmetry of electron changes from four-fold to two-fold \cite{r2,r4}. Meanwhile, symmetry breaking also occurs in lattice and spin degrees of freedom. The lattice undergoes a structural transition from tetragonal to orthorhombic, and a collinear antiferromagnetism transition follows at the same or a lower temperature \cite{r5,r6}. Such intertwining among charge, spin, orbital, and lattice degrees of freedom make it difficult to unveil the primary driving force of the nematic phase transition. In the orbital-driven scenario \cite{r8,r9,r10}, the ferro-orbital order characterized by an occupation difference between the  $d_{xz}$ and $d_{yz}$ orbitals breaks the rotational symmetry and drives the nematic phase transition. In the spin-driven scenario, the nematic phase transition is driven by a spontaneous rotational symmetry breaking of spin fluctuation. The magnitudes of spin fluctuation become inequivalent at (0, $\pi$) and ($\pi$, 0), resulting in a spin nematic phase \cite{r11,r12}. Other scenarios have also been proposed, such as Pomeranchuk instability and complex forms of orbital order \cite{r13,r14,r15}.

Susceptibility describes how strongly a system would respond to an external field. Measuring susceptibility and its temperature dependence provides crucial clues for uncovering the driving force of a phase transition. For the nematic phase transition, the nematic susceptibility has been measured using transport methods \cite{r2,r3}. As it is characterized by the resistivity anisotropy induced by a tunable uniaxial strain, the nematic susceptibility shows a divergent behavior at the nematic phase transition temperature ($T_{nem}$), which indicates a pure electronic origin of the nematic phase. However, the resistivity anisotropy is a macroscopic physical property and thus cannot distinguish how different electronic degrees of freedom play roles in driving the nematic phase transition.

Angle-resolved photoemission spectroscopy (ARPES) is a powerful technique for detecting the electronic structure of materials in the momentum space. Mechanical pressing or stretching devices have been used in ARPES measurements \cite{r23,r24,r25,rdetwin}, which however are not capable of applying a tunable uniaxial strain with high precision, and therefore cannot be used for the measurement of nematic susceptibility. Recently, the piezoelectric stack has been used in ARPES as a strain-tuning device \cite{rpzt}. However, the successful measurement of nematic susceptibly has not been achieved so far.

Here, we show how to conduct a momentum-resolved measurement of nematic susceptibility using a combination of ARPES and an $in$-$situ$ strain-tuning device. The electronic nematic susceptibility is characterized by an energy shift of band induced by a tunable uniaxial strain. We found that the temperature dependence of nematic susceptibility is nonuniform in the momentum space, providing insights into the driving mechanism of the nematic phase transition. Our results demonstrate that the combination of ARPES and $in$-$situ$ parameter tuning is a powerful technique that can measure the electronic susceptibly in the momentum space. Such methodology can be used to study the complex phase transitions in various materials.

High quality FeSe$_{1-x}$S$_x$ single crystals were grown using a chemical vapor transport (CVT) method with a nominal doping level of x~=~0.1. The $T_{nem}$ is around 65~K as confirmed by the resistivity and magnetic susceptibility measurements \cite{r6, r22}. ARPES measurements were performed using a DA30L analyzer and a helium discharging lamp. The photon energy is 21.2~eV. The beam diameter is $\sim$0.5~mm. The overall energy resolution is 8~meV and the angular resolution is 0.3~$^{\circ}$. The typical sample dimension is $\sim$0.7~mm$\times$0.7~mm$\times$0.2~mm. All samples were cleaved $in$-$situ$ and measured in ultrahigh vacuum with a base pressure better than 6$\times10^{-11}$~mbar. Figure 1(a) shows the $in$-$situ$ strain-tuning device used in our experiment. The device is built by ourselves using three lead zirconium titanate (PZT) stacks and titanium blocks to minimize the thermal expansion effect \cite{r20}. An in-plane uniaxial strain can be applied on the sample precisely by changing the PZT voltage \cite{r20, r21}. The PZT expansion ($\Delta$L) was measured $in$-$situ$ using a microscope equipped with a high-resolution camera (see Supplementary Materials \cite{sup} for more details). As shown in Fig. 1(b), the uniaxial strain ($\epsilon$), as characterized by the ratio between $\Delta$L and the sample length (L), increases linearly with the PZT voltage at various temperatures. It is noting that, upon cooling, one need to increase the PZT voltage in order to maintain a constant strain on the sample. No sample deformation was observed during the experiment.

\begin{figure}[t]
\includegraphics[width=8.6cm]{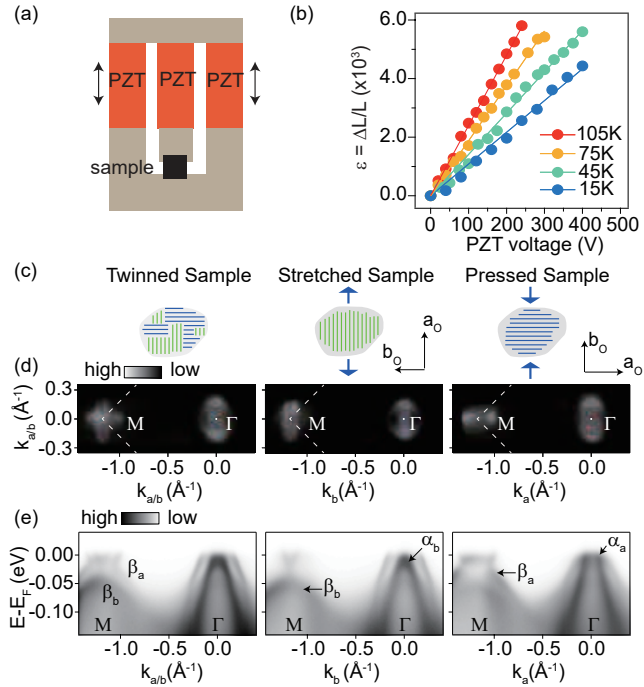}
\caption{Characterization of the $in$-$situ$ strain-tuning device and the sample detwinning effect in FeSe$_{0.9}$S$_{0.1}$. (a) Schematic drawing of the $in$-$situ$ strain-tuning device. (b) The uniaxial strain as a function of PZT voltage at different temperatures. (c) Schematic drawing of the domain orientations in twinned, stretched and pressed samples. $a_o$ and $b_o$ are the longer and shorter unit cell directions in the nematic phase. (d) Fermi surface mappings taken in twinned, stretched and pressed samples. (e) The corresponding energy-momentum cuts taken along the $\Gamma$-M direction. All the photoemission data were taken at 25~K. }\label{f1}
\end{figure}

\begin{figure}[t]
\includegraphics[width=8.6cm]{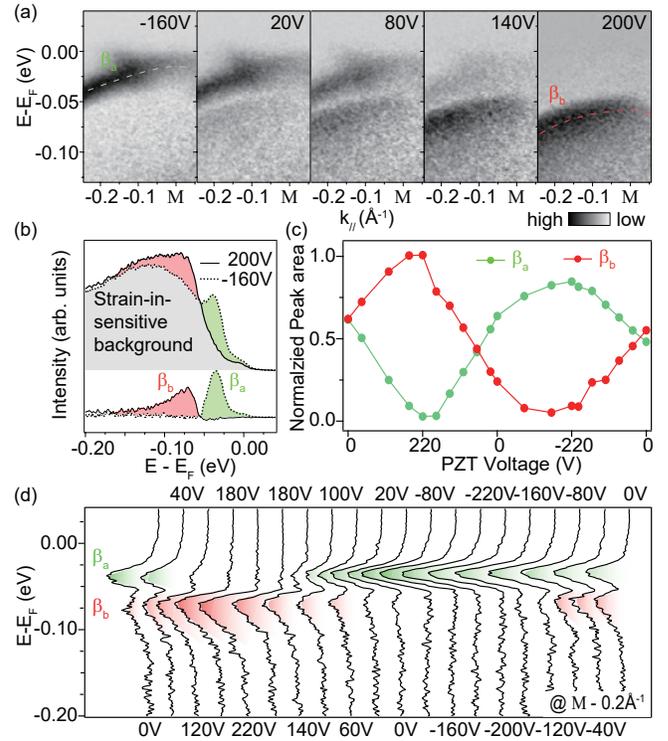}
\caption{Evolution of electronic structure during the detwinning process.  (a) Energy-momentum cuts taken around the M point with different PZT voltages. The spectrum is subtracted by a strain-insensitive background to highlight the spectral weight transfer (see Supplementary Materials  \cite{sup} for more details). (b) Raw and background-subtracted EDCs taken at the momentum (M~-~0.2$\AA^{-1}$). (c) Spectral weights of $\beta_a$ and $\beta_b$ as a function of PZT voltage. The spectral weights of $\beta_a$ and $\beta_b$ are calculated using peak area integrated within [-0.055, 0]~eV and [-0.11, -0.055]~eV respectively. We use unfolded horizontal axis to avoid the controversy induced by the hysteresis effect (see Supplementary Materials  \cite{sup} for more details). (d) Voltage dependence of the background-subtracted EDCs. All data were taken at 25~K. }\label{f2}
\end{figure}

\begin{figure}[t]
\includegraphics[width=8.6cm]{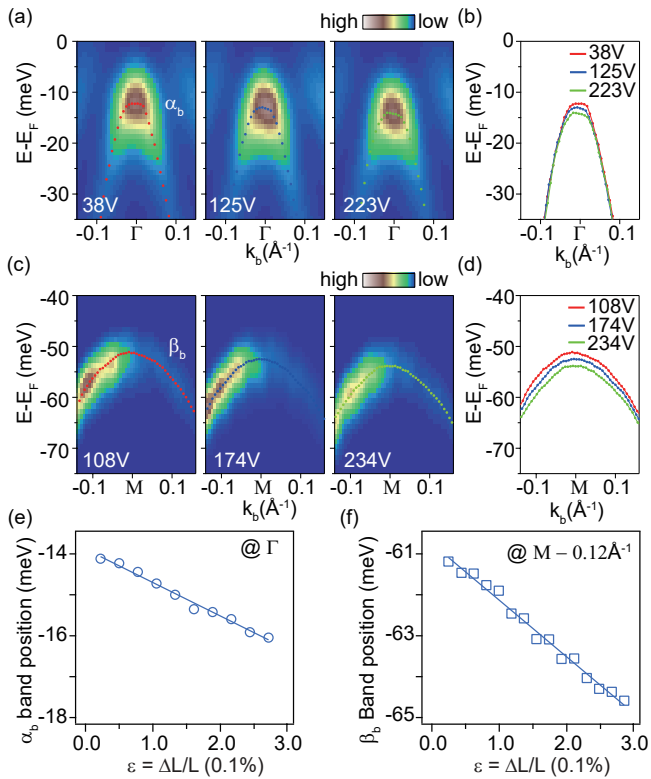}
\caption{The measurement of electronic nematic susceptibility. (a) Energy-momentum cuts taken at the $\Gamma$ point with different PZT voltages. The red, blue, and green points illustrate the band positions obtained by peak fitting. (b) The voltage dependence of the $\alpha_b$ band dispersion. (c) and (d) are the same as panel a and b, but taken on the $\beta_b$ band around the M point. (e) The energy position of the $\alpha_b$ band top as a function of uniaxial strain. The data are fitted using a linear function and plotted in solid line. (f) is the same as panel e, but taken on the $\beta_b$ band. The $\beta_b$ band positions in panel f were taken at the momentum (M - 0.12$\AA^{-1}$). All data were taken at 30~K.}\label{f3}
\end{figure}

\begin{figure*}[t]
\includegraphics[width=16.4cm]{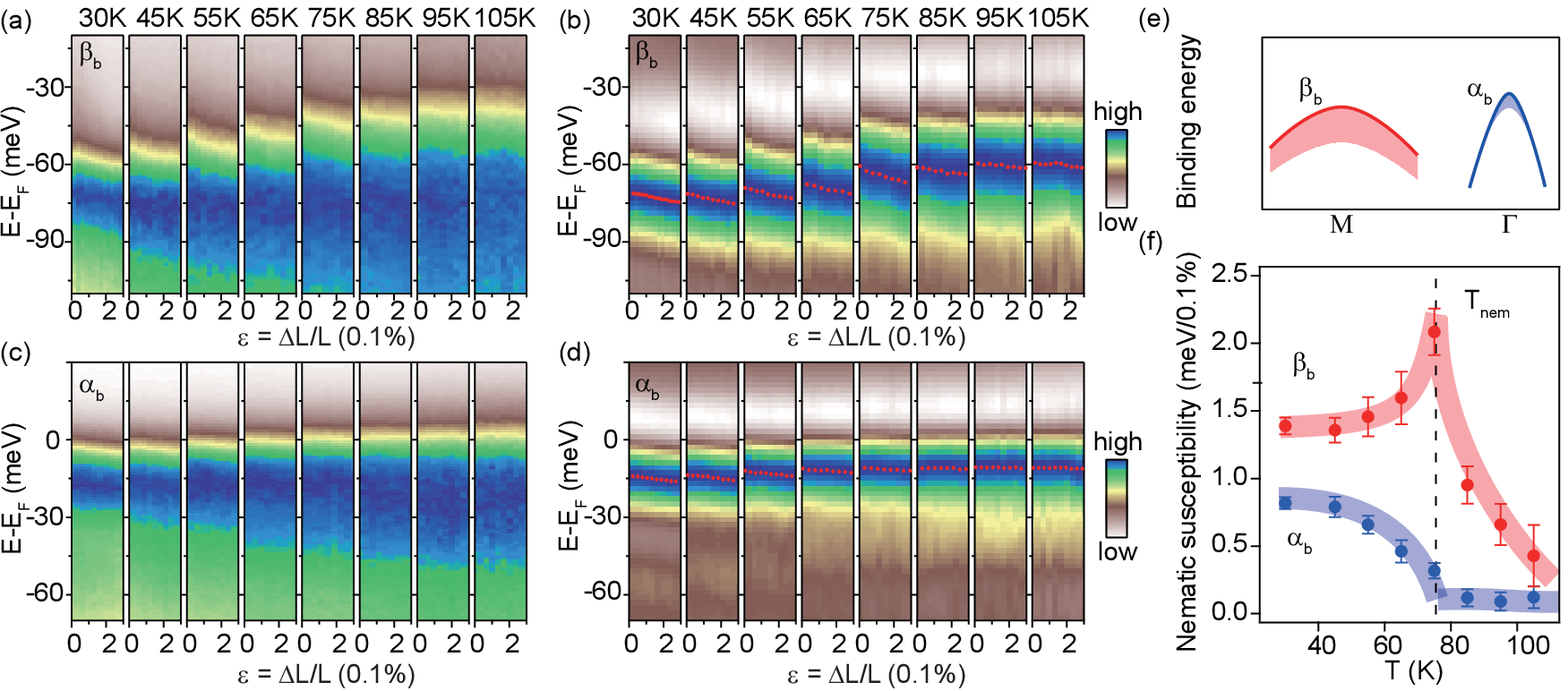}
\caption{The temperature dependence of the nematic susceptibility. (a) The strain dependencies of the $\beta_b$ band position taken at different temperatures. The EDCs taken with different PZT voltages are merged into images to highlight the energy shift of the band position as a function of uniaxial strain. The EDCs were taken at the momenta (M - 0.12$\AA^{-1}$) and $\Gamma$ for the $\beta_b$ and $\alpha_b$ bands, respectively. (b) The second derivative image of the data in panel a. The red points illustrate the band position obtained by the peak fitting of the second derivative EDCs. (c) and (d), are the same as panel a and b, but taken at $\Gamma$ for the $\alpha_b$ band. (e) Illustration of how the $\beta_b$ and $\alpha_b$  bands shift with the increase of nematic order parameter. (f) The temperature dependence of nematic susceptibility take on the $\beta_b$ and $\alpha_b$ bands. The nematic susceptibilities were obtained by linear fitting of the band shifts at different temperatures (see Supplementary Materials \cite{sup} for more details). Error bars were calculated based on the linear fitting process.}\label{f4}
\end{figure*}

When the electronic system enters the nematic phase, two mutually perpendicular directions ($a_o$ and $b_o$) are indiscriminately selected, forming twinned domains in both directions [Fig. 1(c)]. By applying a uniaxial strain, the sample can be detwinned and we can observe the photoemission signal from one single domain \cite{r2, r4, r23,r24,r25}. As shown in Fig. 1(d), the Fermi surface taken in the twinned sample consists of two perpendicularly crossing elliptical pockets at the Brillouin zone corner (M) point. The vertical and horizontal elliptical pockets show up separately in the stretched and pressed samples, indicating that they originate from the vertical and horizontal domains, respectively. Similarly, the band dispersions along the $k_a$ and $k_b$ directions can be probed separately by applying a uniaxial strain. As shown in Fig. 1(e), in the stretched sample, only the lower band ($\beta_b$) can be observed at M, and the band top of $\alpha_b$ is below the Fermi energy ($E_F$) at Brillouin zone center ($\Gamma$). However, in the pressed sample, the upper band ($\beta_a$) emerges at M, and the $\alpha_a$ band crosses $E_F$ at $\Gamma$. All these results are consistent with previous ARPES studies on detwinned FeSe \cite{r23,r24,r25, rdetwin}.

Taking advantages of our $in$-$situ$ strain-tuning device, we can now study how the electronic structure evolves during the detwinning process. Figure 2(a) shows the PZT voltage dependence of the electron bands taken at the M point. When the PZT voltage is -160~V, the sample is pressed and only the upper band ($\beta_a$) could be observed. As the PZT voltage increases gradually from minus to positive, the spectral weight transfers from $\beta_a$ to $\beta_b$. Eventually, only the $\beta_b$ band can be observed in the stretched sample at 200~V. To analyze the data quantitatively, we first choose the energy distribution curves (EDCs) at the momentum (M - 0.2 $\AA^{-1}$) in order to avoid the complexity from the other bands at the M point. Only the $\beta_a$ and $\beta_b$ bands are involved. Second, a strain-insensitive background, as illustrated by the gray area in Fig. 2(b), is subtracted from the EDCs (see Supplementary Materials \cite{sup} for more details). After the background subtraction, the spectral weights of $\beta_a$ and $\beta_b$ can be calculated quantitatively by integrating the peak areas. The result is shown in Figs. 2(c) and 2(d). First, the voltage dependence of the spectral weights clearly shows an anticorrelation between the $\beta_a$ and $\beta_b$ bands, indicating a direct transfer of spectral weight between them [Figs. 2(c) and 2(d)]. Second, the $\beta_a$ and $\beta_b$ bands show similar band dispersions and comparable photoemission intensities. Both facts suggest that the $\beta_a$ and $\beta_b$ bands are two $\beta$ band branches that disperse along the $k_a$ and $k_b$ directions respectively. The spectral weight transfer between $\beta_a$ and $\beta_b$ then reflects the domain rotation during the detwinning process. Our result further shows that the nematic energy splitting characterized by the energy separation between $\beta_a$ and $\beta_b$ is as large as 40~meV at the M point in FeSe$_{0.9}$S$_{0.1}$.

In order to measure the nematic susceptibility, we need to keep the sample detwinned, so that the strain-dependent result would not be complicated by a domain rotation. The nematic susceptibility data were taken on a small sample that could be detwinned easily with a small PZT voltage. Since pressing usually results in a sample deformation, we only focus on the stretched sample where the $\alpha_b$ and $\beta_b$ bands can be observed. As shown in Fig. 3, both the $\alpha_b$ and $\beta_b$ bands shift to higher binding energy with the increase of uniaxial strain. While the whole $\beta_b$ band shifts in a large momentum region, the band shift of $\alpha_b$ only occurs in a narrow momentum region near the $\Gamma$ point. We then plot the band positions at the $\Gamma$ point for the $\alpha_b$ band and at the momentum (M - 0.12 $\AA^{-1}$) for the $\beta_b$ band as a function of uniaxial strain in Figs. 3(e) and 3(f). Both the $\alpha_b$ and $\beta_b$ bands shift linearly with the increase of uniaxial strain. Because the uniaxial strain tunes the nematic order parameter \cite{r2,r3}, the ratio of the band shift ($\Delta$E) to the increase of uniaxial strain($\Delta\epsilon$) can then be taken as the electronic nematic susceptibility of the nematic phase.

With the capability of measuring the nematic susceptibility, we now turn to its temperature dependence on the $\alpha_b$ and $\beta_b$ bands. As shown in Fig. 4, the uniaxial strain increment were kept to be the same for all temperatures. Therefore, the energy shift of bands represents the magnitude of the nematic susceptibility. For the $\beta_b$ band at the M point [Figs. 4(a) and 4(b)], the band shift is observable already at 105~K. Upon cooling, the magnitude of the band shift first increases rapidly, and then decreases when the temperature is below $T_{nem}$. A sharp peak is observed at $T_{nem}$, which is consistent with the transport measurement \cite{r2}, indicating a divergence of nematic susceptibility. However, for the $\alpha_b$ band, the nematic susceptibility only starts to increase when the temperature cools below $T_{nem}$, showing an order-parameter like behavior. It should be noted that, the spin-orbital band splitting is independent with temperature \cite{r27,rsoc}, and therefore does not  affect the temperature dependence of the nematic susceptibility measured at the $\Gamma$ point (see Supplementary Materials \cite{sup} for more details).

As shown in previous ARPES studies, the anisotropic energy shift of bands can be viewed as an order parameter of the nematic phase. However, it is still controversy that such order parameter, that is characterized by the band shift, is nonuniform in the momentum space \cite{r23,r24,r25,rdetwin,rpzt,r27,r16,r26}. For example, the energy scale of the band shift is $\sim$10~meV at $\Gamma$ but $\sim$40~meV at M. The direction of the band shift reverses from $\Gamma$ to M. Here, our measurements indicate that the nematic susceptibility is also nonuniform in the momentum space. If only one order parameter is taken into account, in order to explain the strong momentum dependence of both the band shift and the nematic susceptibility, one need to construct an order parameter that behaves very differently at $\Gamma$ and M. This is inconsistent with all known theoretical models. Alternatively, a more natural way of explaining this controversy is to consider the consistence of two order parameters. One order parameter generates the band reconstruction at $\Gamma$ and the other order parameter is responsible for the band reconstruction at M \cite{r14}.

The coexistence of multiple order parameters has been found in many materials. In some improper multiferroic materials and structure-distorted perovskites, it was found that one primary order parameter could drive the phase transition energetically, while the other secondary order parameters are generated through their coupling to the primary order parameter \cite{rorder1,rorder2}. Here in iron-selenide, both the energy scale of the band shift and the temperature dependence of the nematic susceptibly are different between $\Gamma$ and M. Considering both the large energy scale of the band shift and the divergence of nematic susceptibly at the M point, the order parameter at M is most likely a primary order parameter that energetically drives the nematic phase transition, while the order parameter at $\Gamma$ could be attributed to a secondary order parameter that couples to the nematic symmetry breaking.

We then discuss possible driving mechanisms of the $\Gamma$ and M order parameters. Regrading to the $\Gamma$ order parameter, it has been proposed that the band reconstruction near $\Gamma$ can be well explained by an energy splitting between the $d_{xz}$ and $d_{yz}$ bands, which is consistent with a ferro-orbital order \cite{r24}. However, our results indicate that the ferro-orbital order is likely a secondary order that is only generated below the nematic phase transition due to its coupling to the rotational symmetry breaking. As for the M order parameter, despite of its possible dominating role in driving the nematic phase transition, its microscopic driving mechanism is still unclear. The d-wave or bond orbital order between $d_{xz}$ and $d_{yz}$ have been proposed to explain the band reconstruction observed at the M point\cite{r13, r14}. However, it cannot explain the reconstruction of the $d_{xy}$ bands \cite{r27}. For the spin nematic scenario, it is unclear how the electronic structure would reconstruct when the spin fluctuation becomes highly anisotropic. Further experimental and theoretical studies are required.

The momentum-resolved measurement of electronic nematic susceptibly provides insights to the driving mechanism of the nematic phase in iron-based superconductors. It is intriguing to extend our study to the nematic phase in other materials including cuprates, strontium ruthenium oxides, etc. We can further study how the nematic phase interacts with other intriguing phases. For example, we can directly measure the response of superconducting gap and spin-density-wave gap to the uniaxial strain. Our results highlight the combination of ARPES and $in$-$situ$ parameter tuning as a powerful technique that can disentangle multiple order parameters in the momentum space. This methodology would play an important role in exploring the complex phase transitions in other correlated materials \cite{r28, r29}.

We gratefully thank M. Yi and Donghui Lu for their helpful discussions. This work is supported by National Natural Science Foundation of China (Grant No. 91421107, No. 11574004 and No. 11888101), the National Key Research and Development Program of China (Grant No. 2016YFA0301003 and No. 2018YFA0305602).

\end{document}